\documentclass[]{aa}  
\usepackage{graphicx, amsmath, natbib}
\bibpunct{(}{)}{;}{a}{}{,}
\def\lsim{\mathrel{\rlap{\lower4pt\hbox{\hskip1pt$\sim$}}
    \raise1pt\hbox{$<$}}}                
\begin{document}
   \title{A VLT/FLAMES survey for massive binaries in Westerlund~1: 
\mbox{{\large II.~Dynamical constraints on magnetar progenitor masses from 
the eclipsing binary W13.}}\thanks{This work is
based on observations collected at the European Southern Observatory 
under programme IDs ESO 81.D-0324 and 383.D-0633.}}

   \author{B.~W.~Ritchie\inst{1,2} \and J.S.~Clark\inst{1} 
   \and I.~Negueruela\inst{3} \and N.~Langer\inst{4,5}
}
   
   \offprints{B.W.~Ritchie, \email{b.ritchie@open.ac.uk}}

   \institute{
        Department of Physics and Astronomy, The Open University, Walton Hall, 
        Milton Keynes MK7 6AA, United Kingdom
   \and
        IBM United Kingdom Laboratories, Hursley Park, Winchester, Hampshire 
        SO21 2JN, United Kingdom
   \and 
        Departamento de F\'{\i}sica, Ingenier\'{\i}a de Sistemas y 
        Teor\'{\i}a de la Se\~{n}al, Universidad de Alicante,
        Apdo. 99, 03080 Alicante, Spain
   \and 
        Argelander-Institut f\"ur Astronomie der Universit\"at Bonn, Auf dem 
        H\"ugel 71, 53121 Bonn, Germany
   \and
        Astronomical Institute, Utrecht University, Princetonplein 5, Utrecht, 
        The Netherlands
   }

   \date{Received 21 April 2010/ Accepted 14 June 2010}

   \abstract
  {Westerlund~1 is a young, massive Galactic starburst cluster that contains
   a rich coeval population of Wolf-Rayet stars, hot- and cool-phase 
   transitional supergiants, and a magnetar.}
   {We use spectroscopic and photometric observations of the eclipsing 
   double-lined binary W13 to derive dynamical masses for the two components, 
   in order to determine limits for the progenitor masses of 
   the magnetar \mbox{CXOU J164710.2-455216} and the population of evolved 
   stars in Wd1.}
   {We use eleven epochs of high-resolution VLT/FLAMES spectroscopy to construct 
    a radial velocity curve for W13. $R$-band photometry is used to constrain 
    the inclination of the system.}
   {W13 has an orbital period of 9.2709$\pm$0.0015 days and near-contact 
    configuration. The shallow photometric eclipse rules out an inclination 
    greater than $65^\circ$, leading to lower limits for the masses of the 
    emission-line optical primary and supergiant optical secondary of 
    $21.4\pm2.6$M$_\odot$ and $32.8\pm4.0$M$_\odot$ respectively, rising 
    to $23.2^{+3.3}_{-3.0}$M$_\odot$ and $35.4^{+5.0}_{-4.6}$M$_\odot$ 
    for our best-fit inclination $62^{+3}_{-4}$ degrees. Comparison with 
    theoretical models of Wolf-Rayet binary evolution suggest the emission-line 
    object had an initial mass in excess of $\sim$35M$_\odot$, with the 
    most likely model featuring highly non-conservative late-Case~A/Case~B mass 
    transfer and an initial mass in excess of 40M$_\odot$.}
   {This result confirms the high progenitor mass of the magnetar 
    \mbox{CXOU J164710.2-455216} inferred from its membership in Wd1, and 
    represents the first dynamical constraint on the progenitor mass of any 
    magnetar. The red supergiants in Wd1 must have similar progenitor
    masses to W13 and are therefore amongst the most massive stars to undergo 
    a red supergiant phase, representing a challenge for population models that 
    suggest stars in this mass range end their redwards evolution as yellow 
    hypergiants.}

  \titlerunning{The massive eclipsing binary W13}
  \maketitle
%

\section{Introduction}

The   Galactic   starburst   cluster  Westerlund~1   (hereafter   Wd1;
\citealt{w61}; \citealt{cncg05}) contains  a rich coeval population of
OB supergiants,  yellow hypergiants (YHGs) and  red supergiants (RSGs)
that  collectively map  out the  transitional post-Main  Sequence (MS)
loop redwards followed  by massive stars in the  cluster. The relative
brevity of the transitional phase and intrinsic rarity of such objects
means that  this stage of  evolution is poorly understood,  but recent
downward  revisions to  MS mass  loss rates  \citep{fullerton, mokiem}
suggest that it is of critical importance in understanding how massive
stars shed their outer layers prior to the Wolf-Rayet (WR) phase.

Due to its unique stellar population, Wd1 has been the subject of
intensive observational study in recent years (\citealt{ritchie09a},
hereafter Paper~I, and refs. therein; see also \citealt{clark10};
\citealt{neg10}). Studies of the massive stellar population support a
single burst of star formation at an age $\sim$5Myr and a distance
$\sim$5kpc \citep{crowther06, neg10}, with the identification of
$\sim$O8V stars in the cluster (Clark et al., in prep.) and a
population of lower-luminosity late-O II-III stars just evolving off
the MS both fully consistent with this derived age. Dynamical mass
determinations of late-O dwarfs \citep{gies} and comparison of the
population of OB stars in Wd1 with theoretical isochrones
\citep{meynet00} suggest that stars with
$M_{\text{ini}}$$\sim$30M$_\odot$ lie at the MS turn-off, with the
early-B supergiants having progenitor masses
$M_{\text{ini}}$$\sim$35--40M$_\odot$ \citep{neg10} and the WR
population descended from stars with $M_{\text{ini}}$$\ge$40M$_\odot$
\citep{crowther06}. However, to date no direct mass determination
exists for a member of Wd1; as well as providing confirmation of the
current understanding of the cluster derived from spectroscopic
studies, this is of importance for confirming the high progenitor mass
for the magnetar \object{CXOU J164710.2-455216} that has been inferred
from its membership of Wd1 \citep{muno06}. In addition, the
distribution of evolved stars in Wd1 offers the prospect of demanding
tests for evolutionary models, with both the distribution of WR
subtypes and the large number of cool hypergiants in both YHG and RSG
phases at odds with current predictions \citep{clark10}.

In this paper we present spectroscopic radial velocity (RV)
measurements of the massive binary \object{W13}, identified as a
9.2-day eclipsing system by \cite{bonanos}. X-ray observations reveal
a hard source with $L_x$$\sim$$10^{32}$ergs$^{-1}$, consistent with a
colliding-wind system \citep{clark08}, while subsequent multi-epoch
spectroscopy described in Paper~I showed \object{W13} to be a
double-lined spectroscopic binary consisting of a B0.5Ia$^+$/WNVL
emission-line object\footnote{The WNVL classification is used here to
  indicate strong spectroscopic similarities with other WNL stars in
  Wd1; we note the He~II~$\lambda$4686 line necessary for formal
  classification \citep{crowther97} lies outside our spectral
  coverage.} and an early-B supergiant. We supplement the results of
Paper~I with an additional six epochs of data, providing a total
baseline of approximately 14 months that allows an accurate radial
velocity curve to be derived for the \object{W13} system. This is used
in conjunction with $R$-band photometry \citep{bonanos} to determine
the masses of the two components.


\section{Observations \& data reduction} \label{sec:obs_data}

\begin{table*}
\caption{Journal of observations.}
\label{tab:observations}
\begin{center}
\begin{tabular}{llcc|rr}
Date       & MJD$^{a}$  & Phase$^{b}$ & Elapsed orbits$^{b}$ & RV$_\text{abs}$ (kms$^{-1}$) & RV$_\text{em}$ (kms$^{-1}$) \\
\hline\hline
&&&\\
29/06/2008 & 54646.1846 & 0.33        & 0.33                 & $+$54$\pm9$  & $-$232$\pm7$  \\
18/07/2008 & 54665.0356 & 0.37        & 2.37                 & $+$39$\pm8$  & $-$246$\pm18$ \\
24/07/2008 & 54671.1343 & 0.03        & 3.03                 & $-$50$\pm11$ & -- \\
14/08/2008 & 54692.0423 & 0.28        & 5.28                 & $+$74$\pm13$ & $-$224$\pm14$ \\
04/09/2008 & 54713.0107 & 0.55        & 7.55                 & $-$120$\pm5$ & $+$127$\pm29$$^c$\\
15/09/2008 & 54724.0818 & 0.74        & 8.74                 & $-$200$\pm9$ & $+$163$\pm11$ \\
19/09/2008 & 54728.0554 & 0.17        & 9.17                 & $+$56$\pm4$  & $-$225$\pm17$ \\
25/09/2008 & 54734.0613 & 0.82        & 9.82                 & $-$173$\pm7$ & $+$164$\pm9$  \\
14/05/2009 & 54965.1768 & 0.76        & 34.76                & $-$214$\pm14$ & $+$137$\pm8$  \\
18/05/2009 & 54969.3198 & 0.21        & 35.21                & $+$76$\pm13$  & $-$242$\pm7$  \\
20/08/2009 & 55063.0575 & 0.32        & 45.32                & $+$78$\pm11$  & $-$217$\pm16$ \\
\hline
\hline
\end{tabular}
\end{center}
$^{a}$Modified Julian day at the midpoint of two 600s integrations 
(2$\times$500s, 04/09/2008; 1$\times$600s+1$\times$700s, 19/09/2008).\\
$^{b}$Phase taking $T_0$=54643.080, elapsed orbits for an orbital period of 
9.271d (see Section~\ref{sec:rv}).\\
$^{c}$Unreliable due to the appearance of a central reversal in the emission line,
and excluded from the fit to the RV curve.\\
\end{table*}

\begin{figure}
\begin{center}
\resizebox{\hsize}{!}{\includegraphics{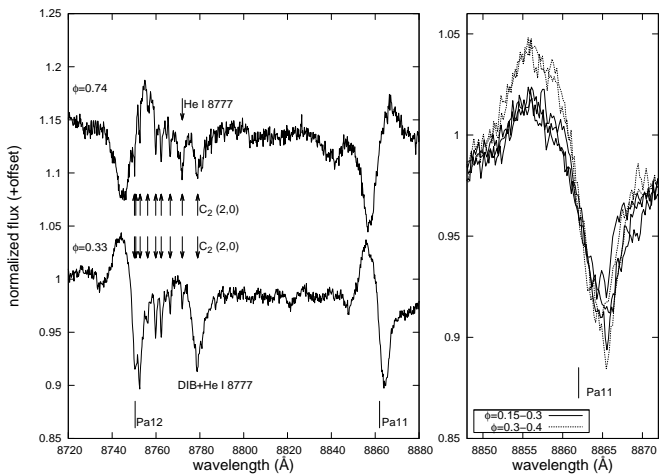}}
\caption{\textit{Left panel}: $I$-band spectra of \object{W13} at $\phi=0.33$ 
and $\phi=0.74$. \textit{Right Panel}: phase dependent variability of the Pa12 emission line. 
Five spectra are overplotted, with three from $\phi=0.15-0.30$ (solid line) and 
two from $\phi=0.30-0.35$ (dotted line).}
\label{fig:phase}
\end{center}
\end{figure}

Spectra of \object{W13} were obtained on eight epochs in 2008 and three epochs 
in 2009 using the Fibre Large Array Multi Element Spectrograph (FLAMES; 
\citealt{pasq02}), located on VLT UT2 \textit{Kueyen} at Cerro Paranal, Chile. 
The GIRAFFE spectrograph was used in MEDUSA mode with setup HR21 to cover the 
8484-9001$\text{\AA}$ range with resolution $R$$\sim$16200; full details of data 
acquisition and reduction are given in Paper~I, and representative 
spectra showing the Pa11~$\lambda$8862 and Pa12~$\lambda$8750 lines at two 
extremes of the RV curve are shown in Figure~\ref{fig:phase}. The signal-to-noise
($S$/$N$) ratio of our spectra is $\sim$95 at 8700$\text{\AA}$. Photometry of 
\object{W13} was taken from the published data of \cite{bonanos}, obtained on 
17 nights between 15/6/2006 and 25/7/2006 using the 1m Swope telescope at Las 
Campanas Observatory, Chile. 

RV measurements were carried out using the IRAF\footnote{IRAF 
is distributed by the National Optical Astronomy Observatories, which are 
operated by the Association of Universities for Research in Astronomy, Inc., 
under cooperative agreement with the National Science Foundation.} 
\textit{onedspec} tasks to fit Gaussian profiles to the cores of the Paschen 
series absorption and emission features in the spectrum of \object{W13}: the 
derived RV at each epoch is an error-weighted average of the RV of individual 
lines. Strong interstellar C$_2$ lines from the (2,0) Phillips band overlap the
Pa12 line, leading to a systematic phase-dependent bias of up to  
$\sim$10kms$^{-1}$ in the line centre, and this line was therefore excluded from 
the analysis. In addition, a broad DIB centred at $\sim$8648$\text{\AA}$ 
\citep{neg10} overlaps the Pa13~$\lambda$8665 line, attenuating the blue flank 
of the emission component at $\phi$$\sim$0.25 and leading to an offset redwards 
relative to the other Paschen series lines, although measurement of the 
absorption component is not affected. Finally, at $\phi$$\sim$0.15--0.3 a
systematic decrease in strength is seen in all Paschen-series emission lines, 
with the lines recovering in strength rapidly after $\phi$$\sim$0.3; this can 
be seen in the right panel of Figure~\ref{fig:phase}, which overplots five 
spectra taken at $\phi$$\sim$0.15--0.35. The weakening in Paschen-series 
emission was assumed to be wind variability in Paper~I, but the persistence of 
this behaviour over more than $\sim$30 orbits in our extended dataset implies 
that it is not a transitory effect but rather a region of excess absorption 
periodically crossing the line of sight.

Absorption line RV measurements were therefore carried using the Pa11
and Pa13--16 lines, noting that the progressive weakening of the
higher Paschen-series lines leads to an increased fitting error and
consequent decreased weight in the derived RV. Emission line RV
measurements were carried out as follows:
\begin{itemize}
\item Emission lines cannot be measured with accuracy for one spectrum
taken near eclipse ($\phi$=0.03, 24/07/2008, MJD=54671.13).
\item At $\phi$=0.1--0.3 attenuation of the blue flank of Pa13 by the 
$\sim$8648$\text{\AA}$ DIB and the decrease in strength of Pa15 and Pa16 leave
only Pa11 and Pa14 emission lines available for measurement, with the two lines 
in good agreement.
\item At other epochs, Pa11 and Pa13-16 emission components were measured. 
\end{itemize}


\begin{figure}
\begin{center}
\resizebox{\hsize}{!}{\includegraphics{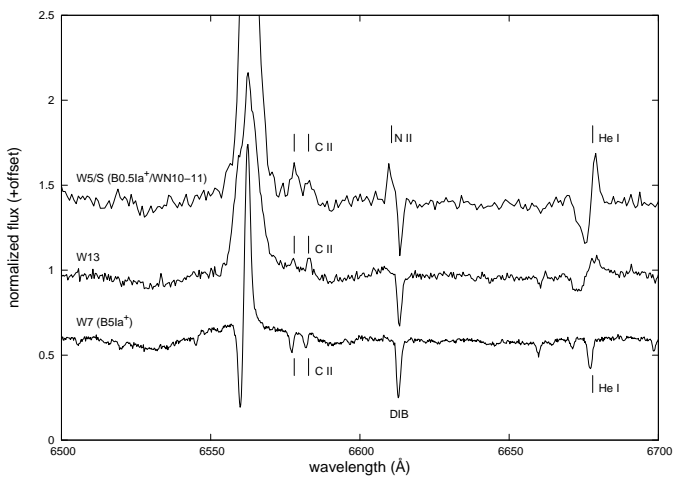}}
\caption{Intermediate-resolution spectra of \object{W13} and the blue 
hypergiants \object{W5} (B0.5Ia$^+$/WN10--11, top) and 
\mbox{\object{W7} (B5Ia$^+$}, bottom). Spectra taken from \cite{clark10} and 
\cite{neg10}.}
\label{fig:compare}
\end{center}
\end{figure}

\section{Results}\label{sec:results}
\subsection{Spectroscopic classification}
\subsubsection{The optical primary}

The $R$-band spectrum of \object{W13} is dominated by the
emission-line optical primary, with strong, relatively broad H$\alpha$
emission (FWHM$\sim$500kms$^{-1}$), complex, time-varying P~Cygni
profiles extending to at least $-350$kms$^{-1}$ in the
He~I~$\lambda\lambda$6678, 7065 lines and weak
C~II~$\lambda\lambda$6578,6582 emission. A comparison of previously
published intermediate-resolution $R$-band spectra of \object{W13} and
the blue hypergiants \object{W5}/\object{WR~S} (B0.5Ia$^+$/WN10--11;
\citealt{neg05}) and W7 (B5Ia$^{+}$; \citealt{neg10}) is shown in
Figure~\ref{fig:compare}. The luminous mid-late~B~hypergiants in
Wd1\footnote{\object{W7}, \object{W33}, and \object{W42a};
  B5--B9Ia$^{+}$ \citep{neg10}.} are all believed to be on a pre-RSG
loop redwards \citep{clark10b}, and display narrow H$\alpha$ profiles
with strong P~Cygni absorption components, along with He~I and C~II in
absorption \citep{clark10, neg10}. In contrast, the spectrum of
\object{W13} shows strong similarities to both \object{W5} and the
WN9h binary \object{W44}/\object{WR~L} \citep{crowther06, clark10},
with the three objects forming a morphologically-distinct group when
compared to the mid-late~B hypergiants, suggesting that the
emission-line object in \object{W13} is also an immediate evolutionary
precursor to the WR phase.

The spectrum plotted in Figure~\ref{fig:compare} 
shows He~I lines that are redshifted by $\sim$50kms$^{-1}$ and C~II emission 
lines somewhat bluewards of He~I (RV$\sim$0kms$^{-1}$), although accurate 
measurement of the C~II line centres is difficult due to their low strength and 
the relatively low $S$/$N$ of the $R$-band spectrum. Nevertheless, if the C~II lines 
originated in the companion then we would expect the lines to be significantly 
blueshifted at this epoch (RV$\sim$$-150$kms$^{-1}$; see Section~\ref{sec:rv}), 
suggesting that the emission lines have a common 
origin\footnote{We note that \object{W5} also appears to show C~II
blueshifted relative to He~I.}. However, 
N~II~$\lambda$6611 emission, present in both \object{W5} and \object{W44}, 
appears absent in \object{W13}, while the H$\alpha$, He~I and C~II emission 
lines are also considerably weaker, suggesting \object{W13} is the least evolved 
member of the WNVL population of Wd1. 

\subsubsection{The optical secondary}

The Paschen-series lines display complex emission/absorption profiles with the 
two components moving in anti-phase (see Figure~\ref{fig:phase}). 
The absorption components have similar strengths 
to the lower-luminosity O9.5Iab/b objects in our FLAMES dataset, but infilling 
from the emission-line object is likely to affect these features and an 
alternative diagnostic is provided by weak He~I~$\lambda$8584, 8777 absorption 
lines that are apparent in many spectra, moving in phase with the Paschen-series
absorption lines. These He~I lines are first seen 
at O8--9I and strengthen rapidly at $\sim$B1.5I \citep{neg10}, with their 
weakness therefore suggesting a spectral type no later than 
$\sim$B1I, although overlapping interstellar features (including a broad, weak DIB at 
$\sim$8779$\text{\AA}$) preclude precise measurement.
A late-O spectral type appears to be excluded by the apparent absence of 
C~III~$\lambda$8500 absorption, with this line leading to a bluewards offset and 
discrepant strength for the C~III~$\lambda$8500/Pa16~$\lambda$8502 blend in the 
O9.5--B0.5 supergiants in Wd1 (\citealt{neg10}; see also Paper~I). 
Neither classifier should be significantly affected by wind emission\footnote{The 
He~I~$\lambda$8584,8777 lines arise 
from transitions from the 3p$^{3}$P$^{\circ}$ level, which is well-populated in 
$\sim$B1--4 supergiants, to upper levels near the ionisation limit of He~I 
\citep{vh72}, but the corresponding downwards transitions are very weak.}, 
and the weak He~I lines and absence of C~III absorption therefore suggests a $\sim$B0.5--1I 
classification with an uncertainty of roughly half a spectral subtype and lack 
of strong constraints on the luminosity class. However, we caution that if C~III 
is weak due to abundance anomalies or near-critical rotation then this limit may 
not apply, and the weakness of the He~I lines permit a classification of 
~O9.5--B0I that is broadly consistent with the Paschen-series line strengths. 

\subsection{Radial velocity curve}
\label{sec:rv}

\begin{figure}
\begin{center}
\resizebox{\hsize}{!}{\includegraphics{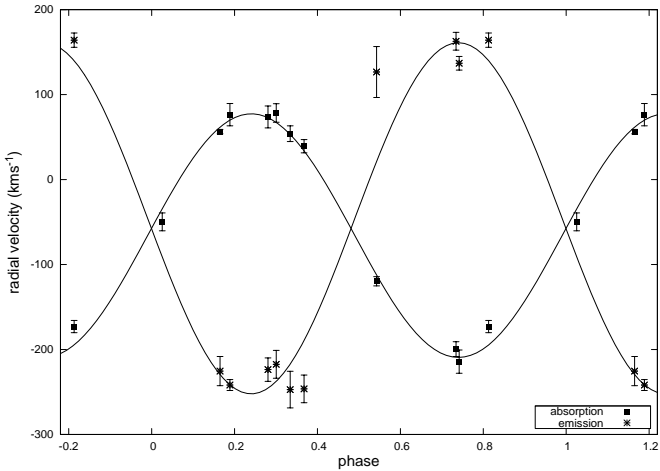}}
\caption{Radial velocity curve for \object{W13}.}
\label{fig:rv}
\end{center}
\end{figure}

Figure~\ref{fig:rv} shows RV curves for the two components of the 
\object{W13} system. Taking the 9.20-day period reported by \cite{bonanos} as a 
starting point, an error-weighted $\chi^2$ fit to the radial velocities of the 
absorption-line component yielded best-fit values for the orbital period of 
\mbox{$P=9.2709\pm0.0015$} days, consistent with an independent determination 
using a Lomb-Scargle periodogram \citep{press}, a systemic velocity of 
\mbox{$-65.9\pm2.4$ kms$^{-1}$} and semi-amplitude 
\mbox{$K_{\text{abs}}=137.3\pm6.7$ kms$^{-1}$}. The corresponding 
fit to the emission line RV curve\footnote{In fitting the emission-line 
RV curve, the discrepant point at $\phi$=$0.55$ was excluded.} has 
a systemic velocity of \mbox{$-48.2\pm3.1$ kms$^{-1}$} and semi-amplitude 
\mbox{$K_{\text{em}}=210.2\pm8.7$ kms$^{-1}$}. Errors are derived from the 
fitting residuals using the bootstrap method \citep{efron}. We note that 
systemic velocity derived from the emission line fit is somewhat lower than that 
derived from the companion, and is in closer agreement with the mean radial 
velocities of other Wd1 supergiants (see Paper~I). Discrepancies in this 
parameter are commonly observed in early-type spectroscopic binaries (e.g. the 
9.8-day O7III(f)+O8.5I binary \object{HD~149404}; \citealt{rauw01}) although the 
effect is small in comparison with some other evolved systems 
(e.g. \object{HDE~228766}; \citealt{massey,rauw02}) in which wind contamination 
strongly affects derived systemic velocities.

Taking these values yields a mass ratio 
\mbox{$q=M_{\text{abs}}/M_{\text{em}}=1.53\pm0.10$} and masses for the two components of:
\begin{equation}
M_{\text{em}~}\text{sin}^3i = \frac{(1+q)^2 PK_{\textsc{abs}}^3}{2\pi G} = 15.9\pm1.9 M_\odot
\end{equation}
and
\begin{equation}
M_{\text{abs}~}\text{sin}^3i = 24.4\pm3.0 M_\odot
\end{equation}

Finally, measurements of blended hydrogen lines with Gaussian fits
tend to yield systematically lower values of $K_1$ and $K_2$ than
methods such as spectral disentangling \citep{simon} that are less
affected by blending \citep{andersen,southworth}. The paucity of
strong lines free from significant interstellar, telluric and wind
contamination in the $R$- and $I$-band spectra of $\sim$B0 supergiants
makes the extent of this effect on our determination of $K_\text{em}$
and $K_\text{abs}$ hard to quantify, and we therefore note that our
method may underestimate the masses of the two components of
\object{W13}.

\subsection{Light curve}

\begin{figure}
\begin{center}
\resizebox{\hsize}{!}{\includegraphics{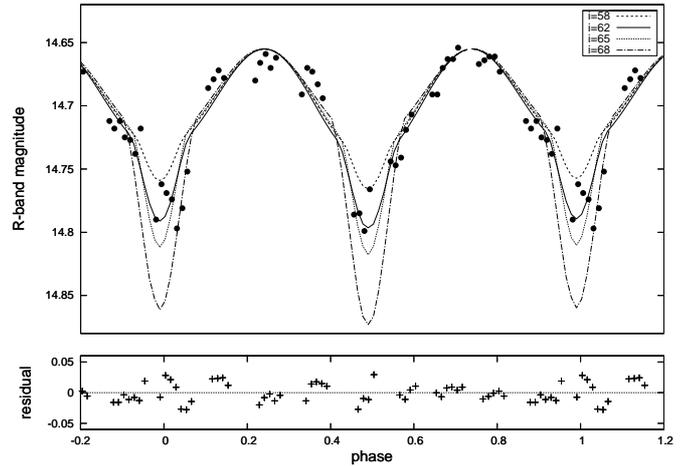}}
\caption{Fit to the R-band lightcurve of W13}
\label{fig:phot}
\end{center}
\end{figure}

To constrain $\text{sin}^3i$, we folded the $R$-band photometric data reported 
by \cite{bonanos} on to the 9.271~day period determined from the RV 
data. The data were binned to reduce the considerable scatter present in the 
light curve, which is most probably a consequence of intrinsic aperiodic 
variability in one or both components: low-level photometric and spectroscopic 
variability is a feature of all transitional supergiants in Wd1, with the blue 
hypergiants displaying rapid photometric variability at the $\sim$0.1 magnitude 
level and the early-B supergiants also variable \citep{bonanos, clark10}. 
Therefore, given the limited dataset and shallow $\sim$0.15 magnitude eclipse, 
we do not expect to be able to fit the light curve of \object{W13} with high 
accuracy. Nevertheless, in the absence of longer-term photometric monitoring, 
the data of \cite{bonanos} allow reasonable constraints to be placed on the 
orbital inclination. 

The \textit{nightfall}
code\footnote{\scriptsize{http://www.hs.uni-hamburg.de/DE/Ins/Per/Wichmann/Nightfall}}
was used to model the light curve of \object{W13}. The effective
temperature of the emission line object was fixed at 25kK, appropriate
for its spectral type, and a linear limb-darkening law and circular
orbits were assumed. The mass ratio was derived from the RV curve, and
the inclination, Roche lobe filling factors for both objects and
temperature of the optical secondary were allowed to vary. The code
rapidly converges to a near-contact configuration in which the
emission-line object has almost filled its Roche lobe (filling factor
$\sim$0.93$\pm$0.05) and the other star has a somewhat lower filling
factor ($\sim$0.74$\pm$0.1).  The best-fit model has an inclination
$i=62^\circ$ and provides a close match to the light curve from
$\phi$$\sim$0.5 to $\phi$$\sim$1, although the region around
$\phi=0.2$--$0.4$ is less well reproduced. Although the
\textit{nightfall} code supports additional features such as `hot
spots' that may provide a better fit to this portion of the light
curve (see, for example, the model of \object{Cyg~OB2\#5} presented by
\citealt{linder}), we consider further refinement of the model
inappropriate given the limitations of the photometric dataset
used. Parameters derived from the light curve model are listed in
Table~\ref{tab:results}, although we stress that our primary goal is
to constrain the inclination of the system and other parameters should
be regarded as provisional pending acquisition of longer-term
photometry.

To examine errors in the derived inclination we investigated models in which $i$ 
is fixed at values from $55^\circ$ to $68^\circ$ while filling factors and 
temperatures are allowed to vary as before; the best fit 
$i=62^\circ$ curve and models with $i=59^\circ$, $i=65^\circ$ and $i=68^\circ$ 
are plotted in Figure~\ref{fig:phot}. Inclinations greater than our best-fit 
model lead to slightly lower filling factors and higher temperatures, 
while the converse is true for lower inclinations. The depth of the eclipse 
provides the strongest constraint on the model, with inclinations greater 
than $\sim$65$^\circ$ strongly disfavoured. $I$-band photometry is also 
presented by \cite{bonanos}, but considerably greater scatter is present in the 
data at mid-eclipse, making it less suitable for modelling. Nevertheless, these 
data also support $i\le65^\circ$, favouring a value $\sim$60--62$^\circ$, although 
the degree of scatter renders this uncertain. Therefore, taking $i\le65^\circ$ places  
robust lower limits of $M_{\text{em}}\ge21.4\pm2.6 $M$_\odot$ and 
$M_{\text{abs}}\ge32.8\pm4.0 $M$_\odot$ for the emission-line object and its 
companion, rising to $23.2^{+3.3}_{-3.0}$M$_\odot$ and 
$35.4^{+5.0}_{-4.6}$M$_\odot$ for our preferred inclination 
$i=62^{+3}_{-4}$$^\circ$. For the purposes of discussion we take the lower 
masses derived from the $i\le65^\circ$ limit. 


\section{Discussion and Conclusions}\label{sec:discussion}

\begin{table}
\caption{Summary of orbital and physical parameters of \object{W13}$^a$.}
\label{tab:results}
\begin{center}
\begin{tabular}{lll}
Parameter & Value\\
\hline\hline
&\\
$T_0$ (MJD)$^a$                       & 54643.080\\
$P$ (days)                            & $9.2709\pm0.0015$\\
$q=m_{\text{abs}}/m_{\text{em}}$      & $1.53\pm0.10$\\
$a$ ($R_\odot$)                       & 72$\pm$3\\
$e$                                   & 0 (fixed)\\
$i$                                   & 62$^{+3}_{-4}$$^{\circ}$\\
\hline
\\
                                      & Emission      & Absorption\\ 
\hline
Filling factor                        & $0.93\pm0.05$ & $0.74\pm0.1$\\
$T_{\text{eff}}$ (K)                  & 25000 (fixed) & 25000$\pm$2000\\
$R$ ($R_\odot$)                       & 22$\pm$2      & 21$\pm$2\\
$\gamma$ (km s$^{-1}$)                 & $-48.2\pm3.1$ & $-65.9\pm2.4$\\
$K$ (km s$^{-1}$)                      & $210.2\pm8.7$ & $137.3\pm6.7$\\
$M$sin$^3i$ (M$_\odot$)               & $15.9\pm1.9$  & $24.4\pm3.0$\\ 
$M (i=65^\circ)$ (M$_\odot$)          & $21.4\pm2.6$  & $32.8\pm4.0$\\
$M (i=62^{+3\circ}_{-4})$ (M$_\odot$) & $23.2^{+3.3}_{-3.0}$ & $35.4^{+5.0}_{-4.6}$\\
\hline
\hline
\end{tabular}
\end{center}
$^a$ Note that $T_0$ corresponds to the eclipse of the B0.5Ia$^+$ emission-line 
star.
\end{table}

\subsection{The evolution of the W13 system}

The short orbital period, near-contact configuration and evolved,
mass-depleted nature of the emission-line object all imply that the
two components of the \object{W13} system must have undergone strong
interaction during their evolution. The 9.3-day orbital period
suggests a late-Case~A or Case~B scenario, with mass transfer
beginning near the onset of shell hydrogen burning
(\citealt{petrovic}, hereafter P05). The presence of unevolved late-O
stars (\mbox{$M_\text{ini}\sim30$M$_\odot$}) in Wd1 suggests a minimum
initial mass $\sim$35M$_\odot$ for shell burning to have commenced,
with in excess of $\sim$10M$_\odot$ lost once mass transfer
begins. However, transfer of angular momentum is expected to lead to
the accretor rapidly reaching critical rotation \citep{packet,
  langer}, while rapid rotation will also greatly increase wind mass
loss rates \citep{langer98}, and a fully-conservative transfer
scenario appears unlikely. Indeed, models of short-period WR+O
binaries by P05 suggests that mass-transfer is highly
\textit{non}-conservative in such scenarios, with only $\sim$10\% of
transferred mass being retained by companion star. The current
21M$_\odot$+33M$_\odot$ (minimum) mass ratio in \object{W13} is
consistent with the P05 model of late-Case~A/Case~B evolution at low
accretion efficiency, with higher accretion efficiencies leading to a
more unequal mass ratio than we observe.

P05 estimate a relationship between initial MS mass and final WR mass
for Case~B systems of \mbox{$M_{\text{ini}} =
  (M_{\text{WR}}+4.2)/0.53$} that suggests that the emission-line
object had an initial mass $\sim$48M$_\odot$.  While this is
consistent with estimates of WR progenitor masses \citep{crowther06},
it is somewhat higher than expected for a star just entering the WR
phase in Wd1. However, \object{W13} will likely shed further mass
before becoming a \textit{bona fide} WN9, suggesting that
$M_{\text{WR}}\sim20$M$_\odot$ and consequent initial MS mass around
$\sim$45M$_\odot$ are more appropriate. Assuming an accretion
efficiency of $\sim$10\% from P05, this would imply that
$\sim$2--3M$_\odot$ was transferred to the secondary, with the
remainder lost from the system. While extended radio emission from
\object{W13} is not detected \citep{dougherty}, emission from both the
O9Ib star \object{W15} and the extreme RSG \object{W26} overlaps the
region around \object{W13} and might obscure direct signs of recent
mass loss.

\object{W13} therefore appears to be a less-evolved analogue to \object{WR21} 
(HD~90657), a 8.3-day 19+37M$_\odot$ WN5 binary considered by P05. One notable 
discrepancy is the presence of an evolved companion in \object{W13}, whereas 
\object{WR21} contains an unevolved mid-O star. Although abundance anomalies  
might suppress the C~III~$\lambda$8500 line in a late-O supergiant, leading to 
an erroneously late spectral type, the presence of He~I lines moving in phase
with the Paschen-series absorption lines excludes a spectral type earlier than 
$\sim$O9 \citep{neg10}. It is possible that the supergiant is still in extreme, 
near-critical rotation from recent mass transfer, and is thus expanded with a 
complex, latitude-dependent spectrum that features both hot (polar) and cooler 
(equatorial) components. However, confirmation of this hypothesis is 
observationally challenging, as infilling of the Paschen-series absorption lines 
prevents direct determination of $v$~sin$i$ while the high reddening 
towards Wd1 precludes the use of fiducial O-type spectral classifiers in 
the blue region of the spectrum.

\subsection{Evolutionary implications for Wd1}
\subsubsection{The cool hypergiant population of Wd1}

\begin{figure}
\begin{center}
\resizebox{\hsize}{!}{\includegraphics{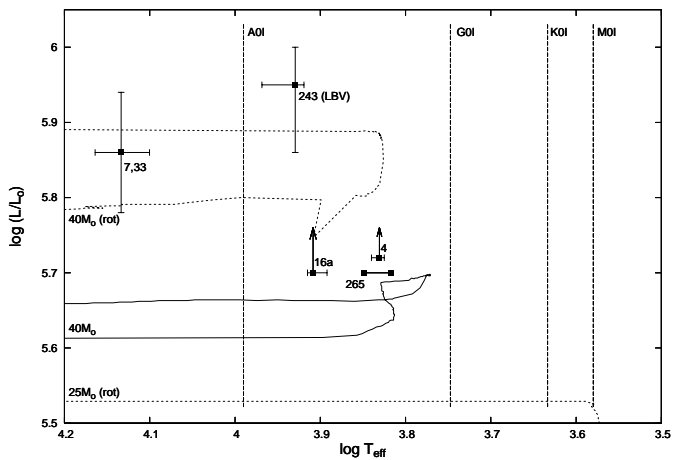}}
\caption{Location of luminous hypergiants in Wd1 for a distance of 5kpc
compared to $z$=0.020 (i.e. solar abundance) evolutionary tracks with 
and without rotation \citep{meynet03}. Lower limits for the
luminosities of the YHGs \object{W4}, \object{W16a} and \object{W265} are plotted. 
\object{W4} is offset vertically by 0.02 dex and error bars are 
omitted for \object{W265} to highlight the change in spectral type over the
pulsational cycle.}
\label{fig:evol}
\end{center}
\end{figure}

The likely $\ge$40M$_\odot$ MS mass for the emission-line object in 
\object{W13}  provides direct constraints on the masses of the eleven cool 
hypergiants in Wd1. Of these, only two have been studied in 
detail\footnote{The A3Ia$^+$ LBV \object{W243} \citep{ritchie09b} and the
YHG \object{W265}, which varies from F1--5Ia$^+$ with a 
$\sim$100~day quasi-period (Paper~I; also \citealt{clark10})}, but the long-term 
dataset compiled by \cite{clark10} reveal a remarkable \textit{lack} of secular 
evolution amongst these objects, with only the LBV \object{W243} 
\citep{ritchie09b} apparently undergoing a major outburst in the half-century 
since the discovery of the cluster by \cite{w61}. 
Although early observations are sparse, the available data are 
sensitive to the long-term evolutionary trends seen in objects such as 
the YHG \object{IRC~+10~420} \citep{hump02}, the LBV \object{R127} 
\citep{walborn} or M33's \object{Var~A} \citep{hump06}. Current observations 
therefore suggest the supergiants in Wd1 undergo a slow redwards 
evolution at approximately constant luminosity until they encounter an extended 
cool-phase state accompanied by growing pulsational instability 
\citep{clark10, clark10b}. 

Evolutionary models of massive stars predict a relatively long YHG lifetime at 
$T\sim6000K$ for stars with $M_{\text{ini}}$$\sim$40M$_\odot$, consistent with 
the lack of secular evolution and mid-A to late-F spectral types for the YHGs in 
Wd1, but do not predict further evolution to the RSG phase \citep{meynet03, drout}. 
However, while the B- and A/F-hypergiant populations plotted in
Figure~\ref{fig:evol} appear in good agreement with the evolutionary 
tracks\footnote{Luminosities for the BHGs \object{W7} and \object{W33}
are taken from \cite{neg10}, while the luminosity of the LBV \object{W243} is
taken from non-LTE modelling assuming a distance of 5kpc \citep{ritchie09b}. YHG
luminosities of \mbox{log$(L/L_\odot)=5.7$} are lower limits from the $M_v$--W(O~I~$\lambda$7774) relationship
\citep{arellano, cncg05}.}, the presence of RSGs in Wd1 suggests that 
stars in this mass range \textit{do} evolve further redwards \citep{clark10}, 
with the extended radio nebulae around these objects \citep{dougherty} 
revealing extensive mass loss that may account for a significant fraction of 
the mass lost prior to the WR phase. Lower-mass stars will not reach the 
RSG phase until $\ge$6Myr, and the absence of any other indicators of 
significant non-coevality in Wd1 \citep{neg10} and the location of two RSGs near 
the core of the cluster argue against the RSGs being descended from a separate
population of older, lower-mass stars. Short-term spectroscopic variability and 
lack of contemporaneous photometry render the luminosities of individual RSGs 
uncertain, and further observations are required to place them firmly on 
the HR diagram. Nevertheless, the M1--5Ia spectral types derived from 
TiO band strengths appear robust \citep{clark10}, and are clearly 
discrepant with respect to current theoretical predictions. 

\subsubsection{The Wd1 magnetar.}

The magnetar \object{CXOU J164710.2-455216} \citep{muno06, muno07} lies 1'.7 
from the centre of Wd1, corresponding to 2.3$(d/5\text{kpc})$~pc in projection. 
The negligible likelihood of the magnetar being a chance association 
and the presence of $\sim$O8V stars in Wd1 provide strong evidence for a massive 
progenitor \citep{muno06}, a result confirmed by our direct measurement of the 
mass of \object{W13} which rules out a magnetar progenitor below 
$\sim$35M$_\odot$ and strongly supports a progenitor mass in excess of 
$\sim$45M$_\odot$ unless mass transfer within \object{W13} is unexpectedly 
conservative. This is consistent with the expected initial masses of the most 
evolved WR stars in Wd1, and also progenitor masses derived for other magnetars: 
a $48^{+20}_{-8}$M$_\odot$ progenitor mass is found for the magnetar 
\object{SGR1806-20} based on its assumed membership of a massive cluster 
at G10.0-0.3 \citep{bibby}, while an expanding HI shell around the magnetar
\object{1E~1048.1-5937} is inferred to be a wind-blown bubble from a 
$\sim$30--40M$_\odot$ progenitor \citep{gaensler}. The progenitor of 
\object{SGR1900+14} has a significantly lower mass \citep{clark08, davies}, 
implying a number of formation pathways exist. However, the derived mass of 
\object{SGR1806-20} is strongly influenced by the distance to the host cluster
obtained from spectroscopic observations, and our results place the first
\textit{dynamical} constraints on a massive magnetar progenitor. Indeed, our 
results suggest that \object{CXOU J164710.2-455216} may have the highest 
dynamically-constrained progenitor mass of \textit{any} confirmed neutron star. 
The high-mass X-ray binary \object{4U1700-37} contains a O6.5Iaf$^+$ mass donor
with mass 58$\pm$11M$_\odot$ but the nature of the 2.44$\pm$0.27M$_\odot$ 
compact object remains uncertain \citep{clark02}, while the 
$\sim$43$\pm$10M$_\odot$, highly-luminous B1Ia$^+$ hypergiant in the X-ray 
binary \object{Wray~977}/\object{GX~302-2} may have evolved via 
quasi-conservative Case~A transfer in a 3.5 day binary with initial masses 
26M$_\odot$+25M$_\odot$ \citep{wellstein, kaper}.

NTT/SofI $K_\text{s}$-band imaging rules out a current $\ge1$M$_\odot$ 
companion to the magnetar \citep{muno06}, a result we confirm using deep 
VLT/NACO $J$-, $H$- and $K_\text{s}$-band imaging (Clark et al., in prep.). 
Nevertheless, given the high binary fraction amongst the WR population 
\citep{crowther06} and the need for a low pre-supernova core mass to avoid 
direct (or fallback) black hole formation (e.g. \citealt{fryer}) it would appear 
likely that the magnetar progenitor \textit{was} part of a (now-disrupted) close 
binary system \citep{clark08}. Support for this hypothesis comes from
population synthesis models, which can only form a neutron star from an isolated 
$\sim$60M$_\odot$ progenitor within the $\sim$5Myr age of Wd1 if mass loss rates 
from stellar winds are greatly enhanced \citep{bel}. In a close binary scenario, 
however, removal of the hydrogen-rich outer mantle via Case~A mass transfer 
results in a reduced post-MS helium core, and ongoing Case~B transfer during 
shell burning will leave a low mass ($\le10$M$_\odot$) helium-burning WR (P05), 
permitting isolated neutron star formation within 5Myr via a type Ib/c supernova 
if the kick velocity is sufficient to disrupt the 
system\footnote{In this scenario, $\sim$90\% of the MS mass of the primary is 
lost prior to SN, with the secondary ultimately forming an isolated black hole 
with $M_{\text{BH}}$$\sim$8M$_\odot$ \citep{bel}.}. Neutron star formation may 
also occur for massive binaries with initial periods greater than a few weeks; 
such systems will not undergo Roche lobe overflow until core hydrogen burning is 
complete and therefore form higher-mass helium cores than Case~A systems, but 
if Case~B or early-Case~C mass transfer can suppress hydrogen shell burning 
before core helium burning is complete then the consequent reduction in the 
mass of the iron core may limit black hole formation \citep{brown}. Further 
results on the distribution of binaries on Wd1 from our VLT/FLAMES survey and 
follow-up observations will therefore allow strong constraints to be placed on 
the formation channels for such systems.
 
\section{Conclusions and future work}

We find lower mass limits for the components of the eclipsing
binary \object{W13} of $21.4\pm2.6$M$_\odot$ and $32.8\pm4.0$M$_\odot$, 
rising to $23.2^{+3.3}_{-3.0}$M$_\odot$ and 
$35.4^{+5.0}_{-4.6}$M$_\odot$ for our best-fit inclination $62^{+3}_{-4}$ degrees,
with spectroscopy suggesting that the evolved emission-line object is likely 
an immediate evolutionary precursor to the WR phase. As conservative mass 
transfer would require the exchange of (at least) 5--10M$_\odot$ without the 
accretor exceeding critical rotation, it appears likely that \object{W13} 
evolved through non-conservative late-Case~A/Case~B mass transfer as the 
(initially) more massive star left the MS. Estimates of the initial WR mass from 
P05 and the presence of a $\ge$33M$_\odot$ supergiant companion, which 
cannot have greatly increased in mass during highly non-conservative transfer,
therefore suggest a MS mass for the emission-line object in excess of 
$\sim$40M$_\odot$. This is consistent with previous estimates of the transitional 
supergiant masses in Wd1 obtained from the MS turnoff and spectroscopy of the 
WR and OB supergiant populations \citep{crowther06, clark10, neg10}. Most 
importantly, this result places the first dynamical 
constraint on the mass of a magnetar progenitor, and highlights a discrepancy
between the presence of RSGs in Wd1 and the predictions of evolutionary models, 
which suggest that the most luminous RSGs should evolve from significantly lower
masses.

A first study of the binary fraction amongst lower-luminosity late-O II-III 
stars in Wd1 will be presented in a subsequent paper in this series, but many 
binary systems are already available for follow-up study. These include 
short-period spectroscopic binaries (\object{W43a}, \object{W3003}; Paper~I), 
eclipsing binaries within the WR, OB supergiant and main sequence populations 
\citep{bonanos}, and X-ray and radio-selected colliding-wind binaries 
\citep{clark08, dougherty}. Consideration of these data will allow further 
dynamical constraints to be placed on the progenitor masses of the evolved stars 
within Wd1 as well as the general mass luminosity relation for stars in the 
upper reaches of the HR diagram and the post-MS pathways they follow. Moreover, 
they will yield the first characterisation of the binary properties of a 
homogeneous population of massive  stars, of critical importance for studies of 
both star and cluster formation and numerous high-energy phenomena such as 
supernovae, Gamma-ray bursters and the formation of high mass X-ray binaries.


\begin{acknowledgements}
We thank Alceste Bonanos for making photometry of Westerlund~1 publicly 
available, Rainer Wichmann for the nightfall code, Paul Crowther for 
comments on a draft of this manuscript, and an anonymous referee 
for a detailed and constructive report.
JSC gratefully acknowledges the support of an RCUK fellowship. IN has been 
funded by grants AYA2008-06166-C03-03 and Consolider-GTC CSD-2006-00070 from 
the Spanish Ministerio de Ciencia e Innovaci\'on (MICINN). 
\end{acknowledgements}


\end{document}